\documentclass[aps,twocolumn,showpacs]{revtex4}
\usepackage{amsmath}
\usepackage{graphicx}
\usepackage{textcomp}
\usepackage{psfrag}

\begin{document}
\title{Correlation between stick-slip frictional sliding and charge transfer}
\author{G. Ananthakrishna}
\author{Jagadish Kumar} 
\affiliation{{Materials Research Centre, Indian Institute of Science, Bangalore 560012, India} }  
\begin{abstract} 

A decade ago, Budakian and Putterman (Phys. Rev. Lett., {\bf 85}, 1000 (2000)) ascribed  friction to the formation of bonds arising from contact charging when a gold tip of a surface force apparatus was dragged on polymethylmethacrylate surface. We propose a stick-slip model that captures the observed correlation between stick-slip events and charge transfer, and the lack of dependence of the scale factor connecting the force jumps and charge transfer on normal load.  Here, stick-slip dynamics arises as  a competition between the visco-elastic and plastic deformation time scales  and that due to the pull speed with contact charging playing a minor role.  Our model provides an alternate basis for explaining most experimental results without ascribing  friction to contact charging.

\end{abstract}

\maketitle
\section{Introduction}
Despite its long history, several aspects of friction remain ill understood even today. This can be partly attributed to the fact that  mechanisms contributing to friction are scale dependent and those operating at lower scales contribute to friction at higher scales. The situation is further complicated as several other factors such as the possible  presence of interfacial layer between the contacting surfaces,  plastic deformation of the contacting asperities, contact electrification etc., are also known to contribute. Recent advances in friction force microscope and surface force apparatus (SFA) have provided useful  tools to understand friction at these scales \cite{Gao04,Hamola90,Carpick96,Berman98}. This coupled with large scale molecular dynamics (MD) simulations \cite{Mo09,Luan06} have provided  a better understanding  of the mechanisms that contribute to friction at nanometer contacts and their relevance to macroscopic friction. Experimental studies show that Amontons' law (the linear relationship between frictional force  and load) holds in a number of situations (from nanometer to macroscopic dimensions) when adhesion is small or the surface is rough (either dynamically induced or otherwise \cite{Gao04,Hamola90,Carpick96,Berman98}). However, smooth adhesive contacts  follow Johnson-Kendal-Roberts (JKR) theory \cite{JKR71,Johnson85}. Recent MD simulations support experimental findings to a large extent \cite{Mo09,Sz08}.

In contrast, the dynamical aspects of intermittent frictional sliding  are less studied \cite{Hamola90,Carpick96,Heslot94}.  Efforts to understand stick-slip dynamics have established that plastic deformation of the contacting interfacial material is responsible for slip \cite{Heslot94,Baumberger99,EB96,Sz08,Landman92,Landman96,SJS96,Chen04}.  A decade ago, Budakian and Putterman \cite{BP00}, proposed a new point of view that friction is due to the formation of bonds arising from contact charging.  This claim however is at variance with  earlier studies that show negligible contribution from contact charging to friction  \cite{Roberts77,Hgraf72,Akande05,Lee94} for charge density $ \sim 10^8$ charges/$mm^2$ reported in Ref. \cite{BP00}.

The authors  demonstrate a correlation between stick-slip events and charge transfer  when the gold tip (radius $0.5$ $mm$) of the cantilever of a SFA was dragged with a velocity of a few $\mu m/s$ on  polymethylmethacrylate (PMMA) surface \cite{BP00}. Typical measured charge density $ \sim 10^8$ charges/$mm^2$.   The magnitude of the slip events is  proportional to the ensuing charge transfer  to the PMMA surface. Interestingly, the total force and the total charge deposited over a scan length collapses onto a single curve when an appropriate choice of a scale factor $\alpha$  is made. The value of $\alpha \sim 0.4$ eV$/{\buildrel _{\circ} \over {\mathrm{A}}}$ is close to the energy window for transfer of charge between the surface states of PMMA and metallic Fermi level \cite{Lee94}.  Further, the authors find $\alpha$ to be nearly constant in the range of normal loads $F_n$  from $68$ to $106$ $mN$. Based on these observations,  they ascribe frictional force to the formation of bonds due to contact charging at the interface. Snapping of bonds leads to slip.  To the best of our knowledge, there is no model that explains these intriguing results.  Our purpose is to construct a model that captures the major results \cite{BP00}. Our model provides an  alternate explanation for the observed correlation and the lack of dependence of the scale  factor $\alpha$ on normal load.  The threshold for slip is determined by plastic deformation of the interface material \cite{Heslot94,EB96} while contact charge plays a minor role.

\section{Background}

\subsection{Motivation and approach}

A notable feature of stick-slip systems is that the stick phase lasts much longer than the slip phase. Then, it is intuitively clear that charge builds up during the stick phase and charge transfer to the PMMA substrate occurs during the slip phase. In view of this, our idea is to first construct a stick-slip model based on contact mechanics of single asperity contact and other relevant physical processes,  and couple it to charging and charge transfer equations. Then, as the contribution from contact charging to friction is small \cite{Roberts77,Hgraf72,Akande05,Lowell80,Lee94},  we expect the correlation between slip events and charge transfer should follow. Indeed, if one is interested in bringing-out the correlation between stick-slip events and charge transfer {\it only}, a 'toy' model for stick-slip that includes velocity weakening law as an input coupled to charging and charge transfer equations is adequate. Such a model has been shown to reproduce the required correlation \cite{Note0}.  However, as our interest is to recover most experimental features, we attempt to include all relevant features of friction and contact mechanics.

Sliding friction is well recognized to be a complex phenomenon with multiple possible scenarios depending on the precise experimental conditions \cite{Gao04,Sz08}.  The nature of the results depend on variety of factors such as the size of the tip radius (typically $10-100 nm$  radius for AFM  and  $mm-cm$ radius for SFA), the load levels, the nature of contacting materials,  the nature of the initial and final microstructures of the tip  and the substrate (smooth or rough), the magnitude of adhesion or the presence of intervening layer etc\cite{Gao04,Sz08,Hamola90,Carpick96,Berman98}. For example, very different conditions such as low adhesion and rough or damaged surfaces can lead to similar friction-load relationship \cite{Gao04,Hamola90,Carpick96,Berman98}. For our case, such detailed information on the precise experimental conditions on the microstructure of the gold tip or the nature of surface of the PMMA substrate before and after the scan, is not provided in  Ref. \cite{BP00}. In view of this and in view of the fact that any model building effort to understand the stick-slip dynamics requires quantitative information on the contact area, we use contact mechanics as the basis of our model. 
Moreover, the JKR theory has been validated in a number of SFA like situations.

\subsection{Single asperity contact, visco-elasticity, plasticity and contact charging}

By necessity our approach to modeling will be dynamical. The idea is to construct a stick-slip model that builds on known results for elastic deformation of the contacting surfaces by adding contributions from visco-elastic and plastic deformations of the contacting surfaces.  Contact mechanics offers a mathematical basis for quantitative description of contact radius and penetration depth. The theory deals with  smooth nonadhesive and adhesive contacts is designed as an equilibrium theory. Yet, experiments show that it is applicable to a number of sliding conditions as well\cite{Gao04,Sz08,Hamola90,CGMS09}.  Non-adhesive contact  was addressed by Hertz later extended to adhesive contacts by Johnson-Kendal-Roberts  and later by Maugis \cite{JKR71,Johnson85,Maugis92}.
In these theories, the area of the contact is a sub-linear function of the normal load $F_n$ \cite{JKR71, Maugis92}. As the Hertz contact area differs less than $5\%$ from the JKR contact area (for normal loads $F_n$ and tip radius $R$ used in \cite{BP00}), we use the Hertz's contact area $A_n$ given by  $A_n = \pi a^2 = \pi [3RF_n/4E^*]^{2/3} = \pi Rz$. Here, $a$ is the contact radius, $z$ is the penetration depth, and $E^*$ is the effective elastic constant (of the contacting materials). The value of $a$ is $\sim 23 \mu m$ for load $F_n =0.1 N$. Thus,   the estimated value of stress is $\sim 60 MPa$ that is higher than the compressive yield stress  of PMMA \cite{Stach97}, but smaller than the yield stress of gold $\sim 80 MPa$. (See Table 1.) Thus, the plastic deformation is largely confined to the weaker PMMA material. This view is supported by MD simulations as well \cite{Sz08,Landman92,Landman96,Chen04,SJS96}.

Quite early, it was shown that accelerated creep of the asperities leads to slip \cite{Heslot94}. Later studies show that stick-slip instabilities arise from two competing mechanisms, namely,  'geometric aging' of the contacting asperities and 'rejuvenation by motion due to plastic deformation' \cite{Baumberger99}. Models for stick-slip instability have been proposed using these ideas \cite{Heslot94,Baumberger99}. Indeed, aging kinetics of contacting asperities has been used   to derive a 'N'-shaped velocity dependent  friction coefficient\cite{EB96}, a generic feature  of most stick-slip systems \cite{Heslot94,Baumberger99,GA07}. Dislocation assisted model for frictional sliding has also been suggested \cite{HK99}.  Further, MD simulations also show  plastic deformation of the contacting interfacial material  and  possible transfer of material between the contacting surfaces \cite{Sz08,Landman92,Landman96,Chen04,SJS96}.  Thus, we consider plastic deformation of the interfacial layer is  responsible for  slip. 

Even though the subject of contact electrification is rather old, the mechanisms underlying the charge transfer between a metal and polymer are still debated \cite{Lowell80,Lee94}. However, the fact that contact electrification contributes to adhesion is well recognized. A number of early studies suggest small contribution to adhesion \cite{Roberts77,Hgraf72,Akande05,Lowell80,Lee94} for the charge levels reported in Ref. \cite{BP00}. An estimate of the  adhesive force from contact charging can be obtained by considering a charged double layer formed at the interface. Noting that contact charging can occur only at the area of contact given by $A_n=\pi a^2 =\pi Rz$, the attractive force is given by $ \pi R z \sigma^2/ 2 \epsilon_0 \kappa$, where $\sigma$ is the charge density and $\epsilon_0$ is the vacuum permittivity  and $\kappa$ the dielectric constant. Using the contact radius ($a \sim 23 \mu m$ for normal load $0.1 N$), the force of attraction is $\sim 10^{-9} N$ that is several orders of magnitude smaller than $mN$ force drops observed in experiments \cite{BP00}. 

Here, it is pertinent to point out that the {\it bond formation attributed  to contact charging} is very different from conventional bonds formed when two atomically 'clean' flat surfaces are brought together or when two surfaces are pressed together. The bonds so formed would have all the relevant electronic contributions (such as kinetic energy of the electron charge density, the long range electrostatic interaction, exchange energy, correlation energy etc). In addition, {\it there could be charge transfer also} when the Fermi energies of the two metals are different or when a metal is in contact with a suitable polymer, as in this case. Indeed, the  process of 'bond formation' is well mimicked by  MD simulations that use appropriate 'potentials' by switching-on the interaction between atoms of the two surfaces when they are brought into contact. 

To summarize, we  list the basic ingredients of our model. First, as stated earlier, we assume that the mechanics of single smooth asperity in contact with a smooth surface is valid for the current situation \cite{JKR71,Johnson85}.  Second, we note that PMMA is a much softer material compared to gold and is a visco-elastic material so that  we include visco-elastic contribution for the deformation. Third,  as shown above, for the load levels and tip radius in the experiment,  stress level can exceed the yield stress of PMMA, but remains less than that of gold.  Thus, we assume that the dissipation is confined to the weaker  PMMA material. (Note however, due to the  mean field nature of our model, our model does not have any scope for dealing such  details as where the dissipation occurs.)   Finally, we include the frictional resistance arising from contact electrification. Thus, in view of the fact that stick-slip dynamics arises from the interplay of all internal relaxational time scales with the applied time scales, we expect that the inclusion of time dependent contributions arising from visco-elasticity and plastic deformation of the contacting interface material will lead to stick-slip dynamics.  

\section{Model Equations}

Our stick-slip model (similar to that of Ref. \cite{VH01})  describes  the center of the contact area $x$ and penetration depth $z$. Our equations are of the general form $m \ddot y = F_a - F_r$, where  $y$  represents $x$ or $z$, $m$ is the mass of the gold tip, $F_a$ is the applied force (shear force $F$ or normal force $F_n$), and $F_r$ is the total  material response.  In equilibrium, $F_a =F_r$. However, in dynamic conditions, there is always an imbalance between the applied force and material response due to time dependent responses from visco-elastic and plastic deformation.  

Consider motion in the $x$ direction.   The reaction force $F_r$ is the sum of the frictional force $F_f$, the adhesive force from contact charging, and  time dependent contributions from  visco-elastic and plastic deformation. Frictional force exerted by the tip on the substrate is $F_f = \tau_0 A_x$, where $\tau_0$ is interfacial shear strength  and $A_x$ is the projected area in the $x-$direction.  The latter is obtained by assuming a parabolic tip (defined by $x^2 = 2Rz$) that gives  $A_x \sim \frac{2}{3} \pi \sqrt{2R}z^{3/2}  = A_{x,0} z^{3/2}$.  In our model, we include both visco-elastic  and plastic deformations of interfacial material.  Visco-elastic effects are  generally included by replacing  the elastic stress $\tau$ by time dependent visco-elastic stress $\tau + \eta \dot \epsilon$, where $\eta$ is the viscosity and $\dot \epsilon$ is the strain rate \cite{NB72,VH01}. Then, the  visco-elastic creep contribution is $\eta_{\parallel}A_x \dot x/D= \eta_{\parallel}A_{x,0} z^{3/2}\dot x/D$, where $\eta_{\parallel}$ is the shear viscosity in the $x-$direction with $D$ a length scale to be fixed. 

The contacting asperities undergo plastic deformation beyond the yield stress $\tau_y$.  The general phenomenological expression for  plastic strain rate  of crystalline materials is given by $ \dot {\epsilon}_p = \dot {\epsilon}_0 exp [ - \frac{\Delta F -\tau {\cal V}_a}{kT}] $, where  $\Delta F$ is the change in the free energy, ${\cal V}_a$ is the activation volume, $\tau$ is the stress, and $\dot \epsilon_0$ is an appropriate prefactor. This expression has been adopted for the case of polymers as well \cite{Stach97}. However, this contains several parameters that are unknown for PMMA. An alternate phenomenological expression for plastic strain rate of crystalline materials that has been used is given by  $ \dot {\epsilon}_p = \dot {\epsilon}_0 (\tau/\tau_y)^n$, where $\tau$ is stress, $\tau_y$ is the yield stress of the material, $n$ an exponent, and $\dot \epsilon_0$ is the strain rate at some specified value of the stress \cite{Hertzberg,Anan04,GA07,Kok03}.  This expression has been successfully used even in the case unstable intermittent plastic flow \cite{GA07,Anan04,Kok03}.   To minimize the number of parameters,  we have adopted this expression for plastic deformation of the interface material. In the physical context, we identify the threshold  stress $\tau_0$ with the yield stress of the softer PMMA material.  Noting that the deformation is continuous beyond the linear visco-elastic flow into the nonlinear plastic flow regime,  the total contribution from  these two flows can be written as $\eta_{\|} A_{x,0}z^{3/2} \frac{\dot x}{D} \Big[ 1- \Big(\frac{F}{A_{x,0}\tau_0 z^{3/2}} \Big)^n\Big]$ \cite{Note}. It is clear that  when $F < A_{x,0}\tau_0 z^{3/2}$, the flow is resistive. This changes over to strain  rate softening flow  when  $F$ exceeds the threshold force $A_{x,0}\tau_0 z^{3/2}$ as the flow rate increases abruptly.  Finally, we include the frictional resistance from contact charging  given by $ \mu A_n \sigma^2/ 2 \epsilon_0 \kappa$ with $\mu$ referring to  the frictional coefficient. We note here that the magnitude of the contact charge contribution (which is $\sim nN$  for the charge level in \cite{BP00}) is several orders of magnitude less than the magnitude of the force drops ( of $mN$ ) in experiments. In contrast, the frictional resistance that is $\sim 1 mN$  as can be verified by using the value of $\tau_0 \sim 10 MPa$ and $z=a^2/R$  in $A_{x,0}\tau_0 z^{3/2}$ (Table \ref{T1}).  However, the frictional resistance from contact charging  can become important when the contact charge density is high. Then,  the inclusion of this contribution  increases the threshold force for slip from $A_{x,0}\tau_0 z^{3/2}$ to $A_{x,0}\tau_0 z^{3/2} + \mu  \pi Rz \sigma^2/2\kappa \epsilon_0$.  
Then, the equations for $x$ and $F$ are
\begin{eqnarray}
\nonumber
\label{tip_eq1}
m \ddot x &=& F -A_{x,0}\tau_0 z^{3/2} -\mu\pi  Rz \frac{\sigma^2}{2\kappa\epsilon_0}  \\
&-&\eta_{\|} A(z) \frac{\dot x}{D} \Big[ 1- 
\Big(\frac{F}{A_{x,0}\tau_0 z^{3/2} + \mu  \pi Rz \sigma^2/2\kappa \epsilon_0}\Big)^n\Big],\\
\dot F &=& K_{\|}(V_a-\dot x),
\label{force1}
\end{eqnarray} 
where $A(z) = A_{x,0}\tau_0 z^{3/2}$.  In Eq. (\ref{tip_eq1}), $F= -K_{\|}(x-V_a t)$ is the applied force with $t$, $K_{\|}$ and $V_a$ referring respectively to time, the effective lateral spring constant and applied velocity. Eq. (\ref{force1}) is the differential form of $F$.  We further make two choices. The first choice is to use  $A(z) = A_0$, a constant $\sim a^2$. Most of the results presented here are for Model I. The second choice is to retain  $A(z) = A_{x,0}z^{3/2}$. We refer to this  as Model II. As we shall show the results of Model II are similar to those of Model I.

Following a similar approach, the equation for $z$  is the difference between the normal force $F_n$  and the upward response of the material $F_r$. This is the sum of elastic contribution and contributions arising from the time dependence of visco-elastic and plastic deformation processes.  The elastic response is given by $\frac{4}{3}R^{1/2} z^{3/2}E^*$. Note that in equilibrium, the balance between the normal load $F_n$ and the elastic response gives  the Hertz relation $a^2= Rz= ( 3RF_n/4 E^*)^{2/3}$. Since, the substrate material is a visco-elastic material, to account for the visco-elastic response, we replace $E^*$ by $E^* + \frac{\eta_{\perp}}{D}\dot z$, where $\eta_{\perp}$ refers to  the bulk viscosity of the PMMA \cite{NB72}. Further, due to the normal load, there can be plastic deformation (of PMMA) when the load and contact area are such the stress $\tau = F_n/ \pi Rz$ is larger than the compressive yield stress $\tau_{y,n}$. Again using the phenomenological relation  $\dot{\epsilon}_p = \dot {\epsilon}_0(\tau/\tau_y)^q$, we can write the plastic strain rate contribution to be  $ -\frac{\eta_{\perp}}{D}\dot z \Big( \frac{F_n}{\tau_{y,n}\pi Rz}\Big)^q $. $q$ is an appropriate exponent for the compressive deformation. Finally, while the tip creeps in the $z$ direction, contact area and hence the position of tip center $x$, also creep in the $x$ direction due to shear force $F$. Noting that $z = z(x(t),t)$, we get an additional creep contribution $ c {\eta}_{\perp} \dot x/D$. Here $ \frac{dz}{dx}= c$ is taken to be a constant. (It must stated that the dependence of $z$ on $x$ is not known. In general  $\frac{dz}{dx}$ would be a function of $z$ which has been assumed to be constant in this case.)  
Combining these contributions, we have    
\begin{eqnarray}
\nonumber
\label{tip_eq2}
\nonumber
m\ddot z &=& F_n- \frac{4}{3}R^{1/2} z^{3/2}\Big[E^*+ \frac{\eta_{\perp}}{D}\dot z \Big ( 1- \big(\frac{F_n}{\tau_{y,n}\pi Rz}\big)^q  \Big) \\
&+ & \frac{c\eta_{\perp}}{D} \dot x\Big] - F \frac{z^{1/2}}{\sqrt{2R}}.
\end{eqnarray}  
The last term is  the normal component of the shear force.  Eqs. (\ref{tip_eq1},\ref{force1},\ref{tip_eq2}) support stick-slip dynamics for a range of parameter values. 

In our model, stick-slip instability arises due to a feed back loop. To see this, we note that when $\dot x$ is small (stuck phase), $F$ increases (see Eq. (\ref{force1})). Concomitantly, $z$  approaches the equilibrium value at a rate controlled by $\eta_{\perp}$. Thus, the frictional resistance $\tau_0A_{x,0}z^{3/2}$ and contact charging resistance $\mu\pi Rz \frac{\sigma^2}{2\kappa\epsilon_0}$ increase. Since $\dot x < V_a$, $F$ keeps increasing till the threshold for slip is crossed triggering an abrupt increase in $\dot x$  due to plastic flow (fifth term in Eq. (\ref{tip_eq1})).  This causes the force to drop as $\dot x$ overshoots beyond $V_a$ during the slip duration (see Eq. (\ref{force1})). The whole process starts all over again since the force has dropped below the threshold for slip. 

Assuming a finite time scale for charging after contact is established between the tip and PMMA, the  equation for the charge density $\sigma$ at the contacting surface  is 
\begin{equation}
\dot \sigma = {\sigma_m \over t_a}\Big(1-{\sigma \over \sigma_m}\Big) -{\frac{\dot x}{D} \sigma}, 
\label{charge1}
\end{equation}
where $\sigma_m$ is the saturation value and $t_a$ a time constant. 
Once slip occurs, contact charge is transfered to the PMMA surface at a local slip rate denoted by the last term in Eq. (\ref{charge1}).  Using Eq. (\ref{charge1}) and the fact that charging can only occur at the area of contact determined by the load,  we get the equation for the total charge $\sigma_t = \pi R z \sigma$,  
\begin{equation}
\dot \sigma_t = \frac{(\pi R z \sigma_m - \sigma_t )}{t_a} + \sigma_t {\frac{\dot z}{z} - \dot x \frac{\sigma_t}{D} }. 
\label{charge2}
\end{equation}
The last term is the charge transferred to the PMMA surface during a slip event. Then, the charge transferred during each slip event is given by $\dot \sigma_d = \dot x \sigma_t/D$.

\subsection{Scaled Equations}

We cast the equations for Model I (i.e., $A(z) = A_0$) in a dimensionless form using  a basic length scale $D = F_{max}/K_{\|}$ and a time scale determined by $\omega^2=K_{\|}/m$.  Then, the scaled variables are : $x=XD$, $z=ZD$, $F^s=F/K_{\|}D$, $\bar{\tau}_0=\tau_0 A_{x,0} D^{1/2} /K_{\|}$,
$\bar{F}_n=F_n/K_{\|} D$, the normal spring constant $K_{\perp} = \frac{4}{3} (RD)^{1/2} E^*$,  $\bar{\eta}_{\perp}= \frac{4}{3} (RD)^{1/2} \omega \eta_{\perp}/K_{\|}$, $\bar{ \eta}_{\|}=A_0 \eta_{\|}\omega /K_{\|}D$,  ${\bar \tau}_y=\tau_{y,n} \pi R/K_{\|}$, $v_a = V_a/\omega D$,  and  $T_a=t_a \omega$. Defining the scaled total charge $\Sigma=  \pi RD Z\sigma/\sigma_0$ with $\sigma_0 =( 2\pi \kappa \epsilon_0 RD^2K_{\|})^{1/2}$ and  $\Sigma_m= \pi RD\sigma_m/\sigma_0$,  the dimensionless equations are:
\begin{eqnarray}
\label{X-eqn}
\ddot X &=& F^s-\bar{\tau}_0Z^{3/2} -\mu \frac{\Sigma^2}{ Z} -\bar{\eta}_{\|}\dot X\Big[ 1- 
\Big(\frac{F^s}{\bar{\tau}_0Z^{3/2} + \mu \frac{\Sigma^2}{ Z}}\Big)^n\Big], \\
\nonumber
\label{Z-eqn}
\ddot Z &=& \bar{F_n}- \Big[\frac{K_{\perp}}{K_{\|}} + \bar{\eta}_{\perp}\dot Z \Big( 1- \Big(\frac{ {\bar F}_n}{\bar{\tau}_y Z}\Big)^q\Big)\\
&+& c\bar{\eta}_{\perp}\dot X \Big]Z^{3/2} - \sqrt{\frac{D}{2R}}F^s Z^{1/2}\\
\label{S-eqn}
\dot \Sigma &=& {\Sigma_m Z-\Sigma \over T_a} -{\dot X \Sigma}+\frac{ \dot Z}{Z} \Sigma, \\
\label{F-eqn}
\dot F^s &=& v_a-\dot X.
\end{eqnarray}
The scaled equation for charge transferred (to the substrate) is $\dot \Sigma_d = \dot X  \Sigma$.  As there is only one way coupling between $(X,Z)$ variables and $\Sigma$ and $\Sigma_d$, the instability domain of Eqs. (\ref{X-eqn},\ref{Z-eqn},\ref{S-eqn},\ref{F-eqn}) are nearly the same as the stick-slip model (Eqs. (\ref{tip_eq1},\ref{force1}, \ref{tip_eq2})) for $\sigma_m \sim 10^{-5}$ $C/m^2$ reported in Ref. \cite{BP00}. However, when the charge levels are much higher, the threshold for slip increases and  the instability domain is altered from that at low levels of charge.  For Model II, the scaled equation for $X$ is given by 
\begin{equation}
\ddot X = F^s-\bar{\tau}_0Z^{3/2} -\mu \frac{\Sigma^2}{ Z} -{\eta}^{\prime}_{\|}Z^{3/2}\dot X\Big[ 1- 
\Big(\frac{F^s}{\bar{\tau}_0Z^{3/2} + \mu \frac{\Sigma^2}{ Z}}\Big)\Big],
\label{X-eqn1}
\end{equation}
where ${\eta}^{\prime}_{\|} = a_{x,0} D^{1/2} \omega \eta_{\|}/K_{\|}$. Note that except for ${\eta}^{\prime}_{\|}$, all other parameters are the same. Most numerical results are for $n=q=1$. For the sake of completeness, we have presented results for $n=1.75,q=2$ also.

Table \ref{T1} shows the parameter values (for Model I). Values in the  second row are from \cite{BP00}. As no material parameters are given in \cite{BP00}, values of unscaled parameters (fourth row)  are taken from the literature wherever available. In particular, viscosity $\eta$ that depends on molecular weight, temperature, shear rates etc, is not available.  The magnitude of $\tau_0 \sim \mu \tau_s$, where $\tau_s$ is the shear yield stress and $\mu \sim 0.3$ is the friction coefficient.  Yield stress depends on strain rate \cite{Stach97}.  In addition, (for PMMA valid also for  polymers in general) compressive yield stress is always higher than the shear yield stress \cite{Stach97}. The range of values  of these quantities are listed in the fourth row. The scaled parameters used for the calculations are in the last row.  Other parameters used are: $A_0\sim a^2\sim 10^{-10}m, \kappa= 3.3, v_a=1.45\times 10^{-5}, T_a=2$ and $c= 0.122$. Choosing $F_{max} =47$ $mN$ gives $D =10^{-3} m$ and thus the range of $\bar{F}_n$ corresponding to $68-106mN$ is $1.446-2.26$. 
\begin{table}[!h]
\caption{Parameter values used for the model. Values in the second row are from Ref. \cite{BP00}. See text.}
\label{T1}
\begin{center}
\begin{tabular} {|c |c |c |c |c |c |}
\hline 
 $R$ & $m$ &$K_{\|}$& $F_n$& $V_a$&$\sigma_m$\\
(mm) &($kg$)& ($N/m$)&($mN$)&($\mu m/s$)&($C/m^2$) \\
\hline
$0.5$  & $10^{-5}$ & $47$ & $68-106$ & $ \sim 10 $  & $1.67\times 10^{-5}$   \\ [0.5ex]
\hline
$E^*$ & ${\eta}_{\|}$ & ${\eta}_{\perp}$ & $ {\tau}_0$ & ${\tau}_{y,n}$ & $\sigma_0$  \\

GPa & Pa.s &  Pa.s & MPa &  MPa & ($nC/m^2$) \\
\hline
$1 - 3$  & $...$& $...$ & $0.1-10$ & $1-50$  & $3.0$   \\ [0.1ex]
\hline
$K_{\perp}$ & $\bar{\eta}_{\|}$ & $\bar{\eta}_{\perp}$ & $\bar {\tau}_0$ & $\bar {\tau}_y$ & $\Sigma_m$  \\ 
\hline
$\sim 10^6$  & $10^{2}$ & $8\times 10^8$ & $1.0$ & $2500 $  & $0.0167$   \\ [0.5ex]
\hline
\end{tabular}
\end{center}
\end{table}

\begin{figure}[!h]
\vbox{
\includegraphics[height=5cm,width=7.5cm]{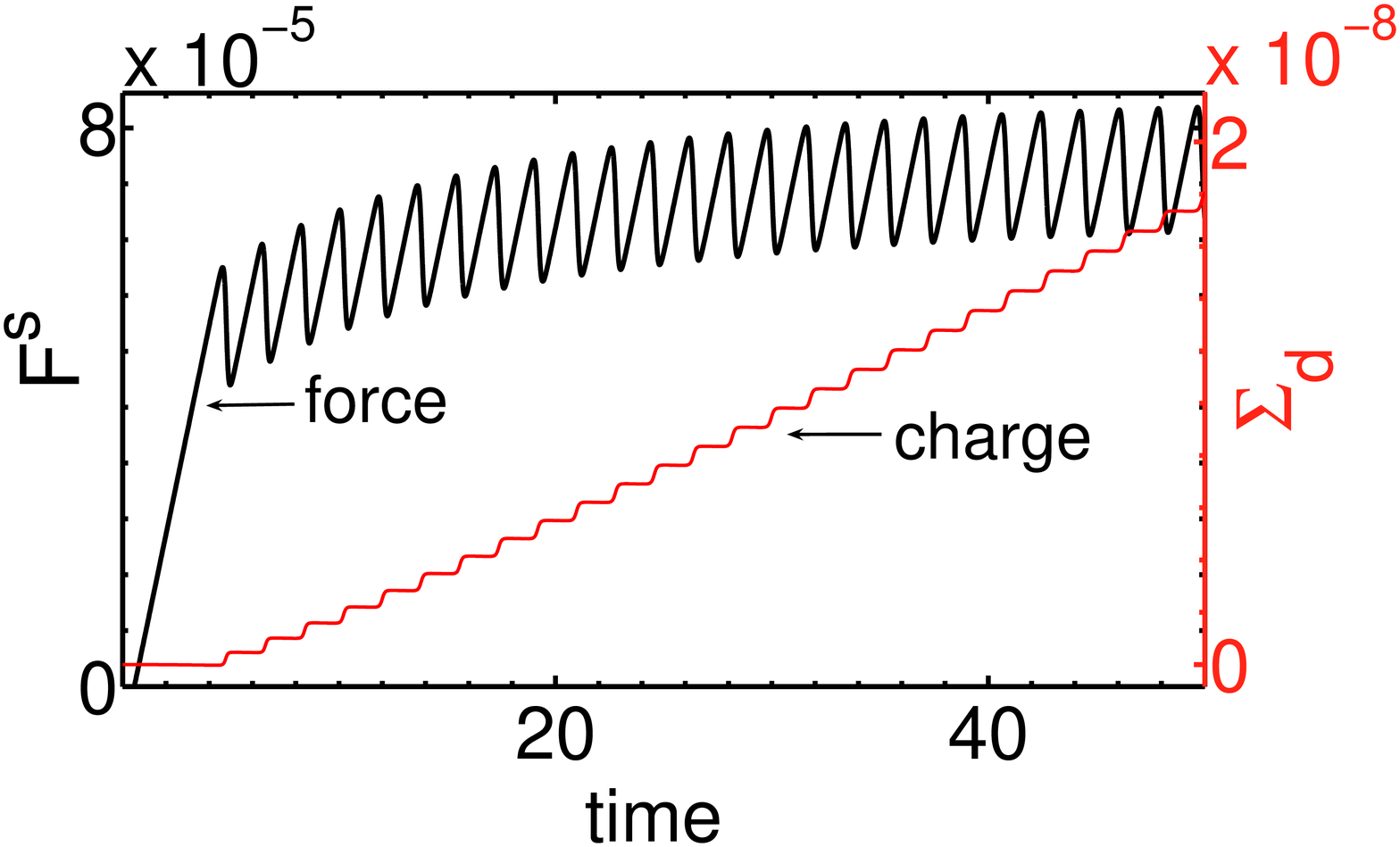}
\includegraphics[height=5cm,width=7.2cm]{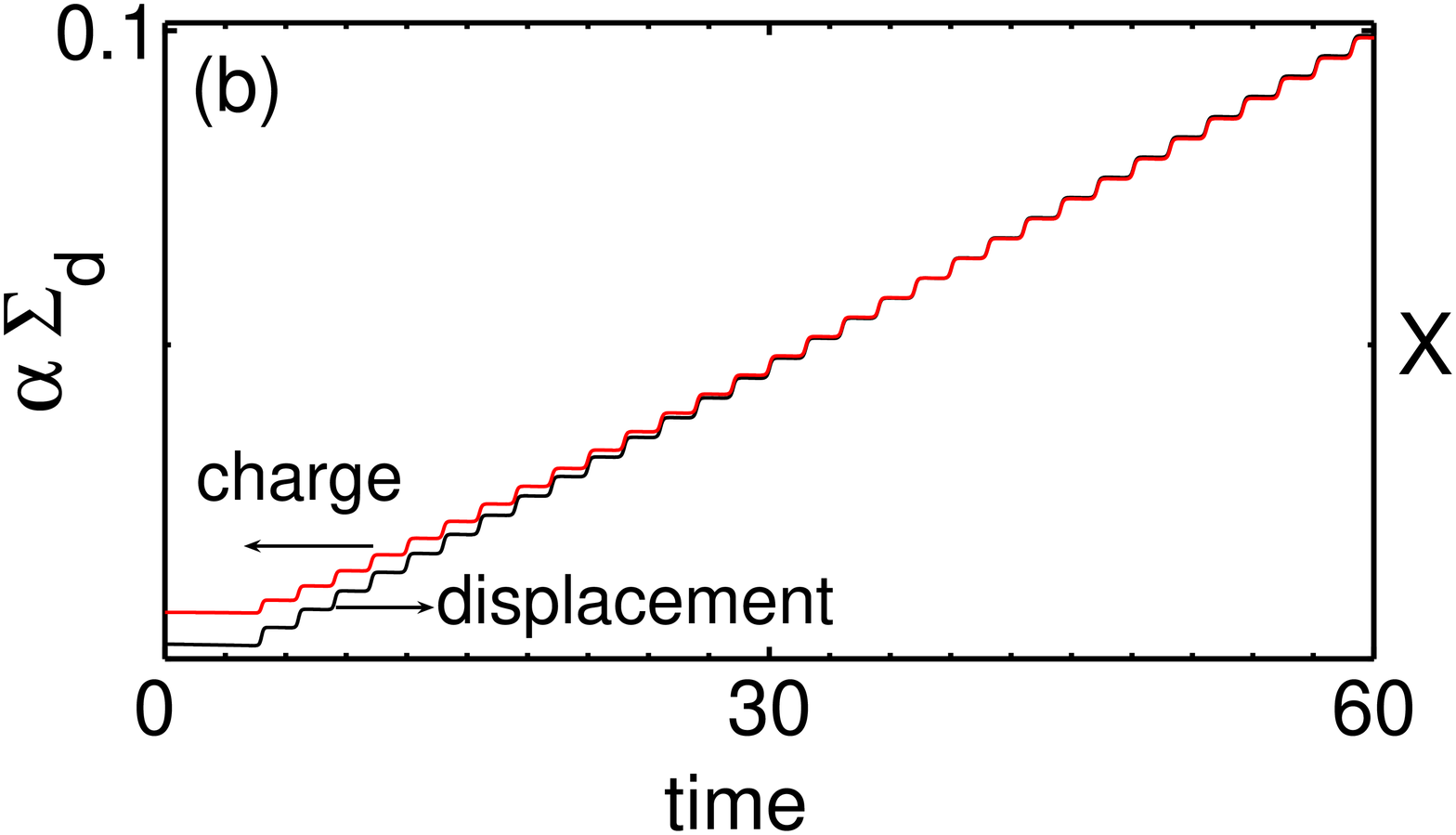}
}
\caption{(color online) (a) Plots of scaled force and cumulative charge transferred to the substrate. (b) Collapse of the mean displacement $X$ and charge transferred $\Sigma_d$ for $\bar{F}_n=2.0$ for a scale factor $\alpha_s = 3.37 \times 10^4$. 
}
\label{FPM}
\end{figure} 

\section{Results}

We first examine the stability of the uniform sliding state for Model I (Eqs. (\ref{X-eqn},\ref{Z-eqn},\ref{S-eqn},\ref{F-eqn})) or for Model II (Eqs. (\ref{X-eqn1},\ref{Z-eqn},\ref{S-eqn},\ref{F-eqn})).  Here also we use $n=q=1$ for simplicity in calculation.   As stated earlier, for charge levels of $\sigma \sim 10^{-5} C/m^2$, the contribution from the third term in Eq. (\ref{X-eqn}) (or Eq. (\ref{X-eqn1})) is small compared to the frictional term. (See Table \ref{T1} and note that $\Sigma$ approaches its saturation value given by  $ \Sigma_m Z_{asy}$, where $Z_{asy}$ is the near saturation value of $Z$ determined by the balance between $\bar{F}_n$ and $K_{\perp} Z^{3/2}/K_{\|}$.). Thus, we can conveniently examine the stability of Eqs. (\ref{X-eqn},\ref{Z-eqn},\ref{F-eqn}) (or Eqs. (\ref{X-eqn1},\ref{Z-eqn},\ref{F-eqn}) ) by dropping the charge contribution. A standard way  is to carry-out the stability analysis of these equations in the steady state sliding conditions (i.e.,  $\Dot X = v_a, \ddot X = 0, \dot Z= 0$ and $\ddot Z=0$ valid in most piezo controlled SFA situations) is to find the eigen values of the linearized system of equations. However, it turns out that even the calculation of fixed points is messy and the expressions are complicated, forcing us to carry out the analysis numerically (not presented here). Instead, a simpler method is find $F^s$ as a function of the pull velocity \cite{VH01} in the steady state. Setting $\ddot X= 0$ in Eq. (\ref{X-eqn}) (or \ref{X-eqn1}),  both Model I and II give $F^s = \bar{\tau}_0 Z^{3/2}$. Setting $\ddot Z= \dot Z= 0$, we get
\begin{equation}
Z^{3/2} = \frac{{\bar F}_n}{[\frac{K_{\perp}}{K_{\|}} + c\bar{\eta}_{\perp} v_a]},
\end{equation}
where to simplify the calculations,  we have dropped the term $\sqrt{\frac{D}{2R}}F^s Z^{1/2}$ in Eq. (\ref{Z-eqn}) as this term is small compared to $\frac{K_{\perp}}{K_{\|}} Z^{3/2}$. Using this in $F^s = \bar{\tau}_0 Z^{3/2}$, we get
\begin{equation}
F^s = \frac{ \bar{\tau}_0{\bar F}_n}{[\frac{K_{\perp}}{K_{\|}} + c\bar{\eta}_{\perp} v_a]}.
\end{equation}
Clearly $\frac{dF^s}{dv_a}$ is negative even from very small values of the pull velocity $v_a$. Thus, the steady sliding state is unstable.  However,  there is a lower cut off in $v_a$ that can be calculated once  $\sqrt{\frac{D}{2R}}F^s Z^{1/2}$ is included. This ensures stick-slip oscillations for the systems of equations. Note that the static friction threshold is obtained by setting $\dot X = v_a=0$ which gives $F^s = \frac{{\bar \tau}_0{\bar F}_n K_{\|}}{K_{\perp}}$. The analysis also shows that the static friction is larger than the sliding friction.

We now consider the numerical solution of Eqs. (\ref{X-eqn},\ref{Z-eqn},\ref{S-eqn},\ref{F-eqn}). The force-displacement curve is shown in Fig. \ref{FPM}(a) for $\bar{F}_n=2.0$ ($F_n=94$ $mN$) along with the cumulative charge transferred to the substrate $\Sigma_d$. Clearly, the mean force $F^s$ gradually increases and saturates, a feature seen in experiments (Fig. 3(a) of Ref. \cite{BP00}).  Further, the average slope of $\Sigma_d$ as a function of time (in the asymptotic regime) is proportional to that for the displacement $X$. Using a proper scale factor $\alpha_s$ these two curves can be made to collapse onto a single curve shown in  Fig. \ref{FPM} (b) for $\alpha_s  = 3.37 \times 10^4$.  Thus, the model recovers the correlation between the stick-slip events and charge transfer.

We now examine the influence of the normal force $\bar{F}_n$.  We have calculated displacement $X$ and cumulative charge transferred  $\Sigma_d$ for ${\bar F}_n = 1.446 -2.26$ ($F_n=68-106$ $mN$). In each case,  $X$ and $\Sigma_d$ curves collapse onto a single curve for a proper choice of the scale factor $\alpha_s$.  In particular, for ${\bar F}_n = 1.446$ and $2.26$ the values of $\alpha_s$ respectively are  $4.18 \times 10^4$ and $ 3.09 \times 10^4$. Then, the ratio of the maximum deviation ($1.09 \times 10^4$) to the mean ($3.635\times 10^4$), is $\sim 30\%$. This compares well with $25\%$ scatter in $\alpha$ in experiments for $F_n = 68-106$ $mN$ (Fig. 3(b) inset of \cite{BP00}). This shows that  the change in $\alpha^s$ (equivalently $\alpha$) is small for the limited range of normal loads considered ($\bar{F}_n = 1.446-2.26$ or $F_n = 68-106mN$). 

\begin{figure}[!h]
\vbox{
\includegraphics[height=5.0cm,width=7.5cm]{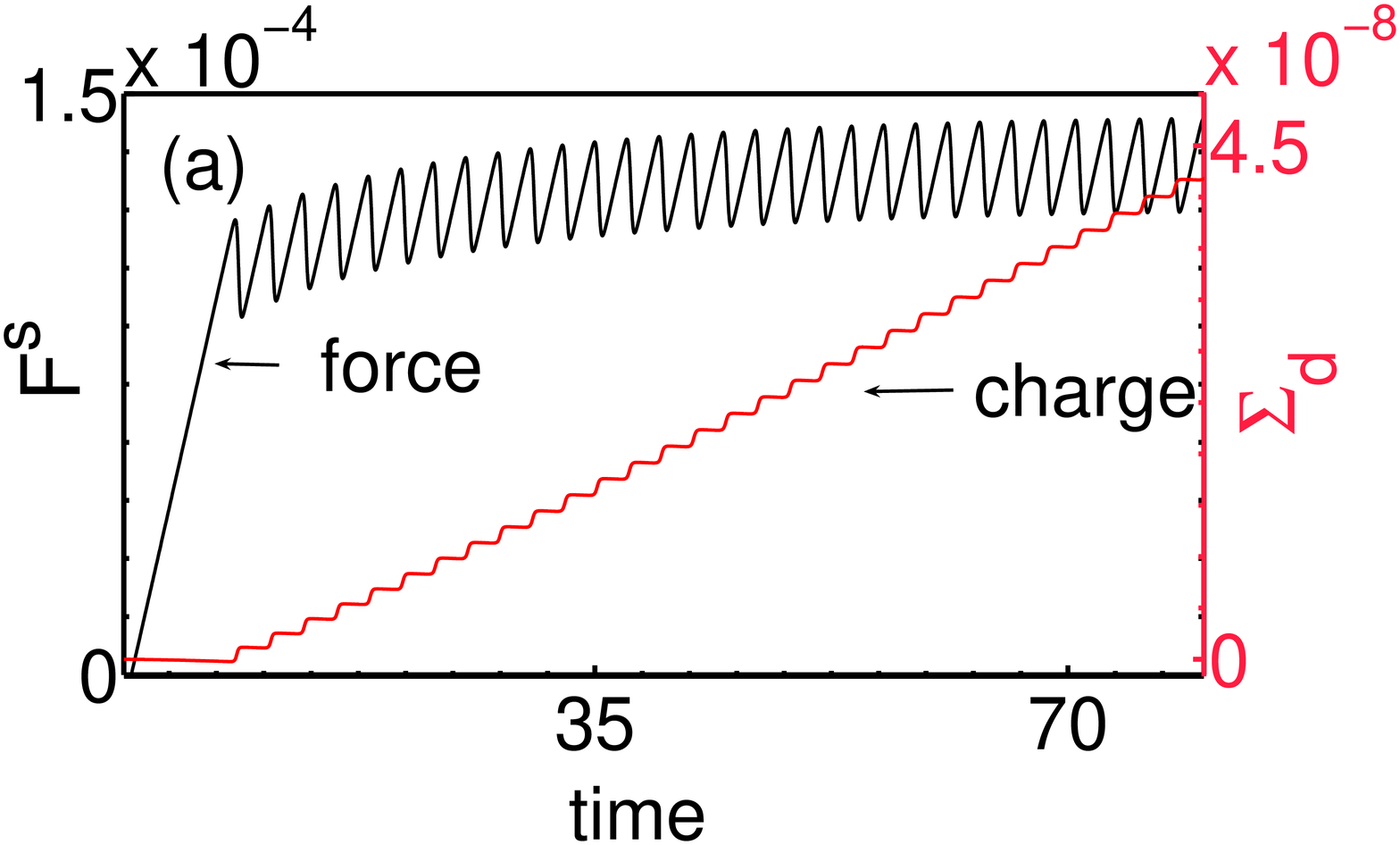}
\includegraphics[height=5.0cm,width=7.2cm]{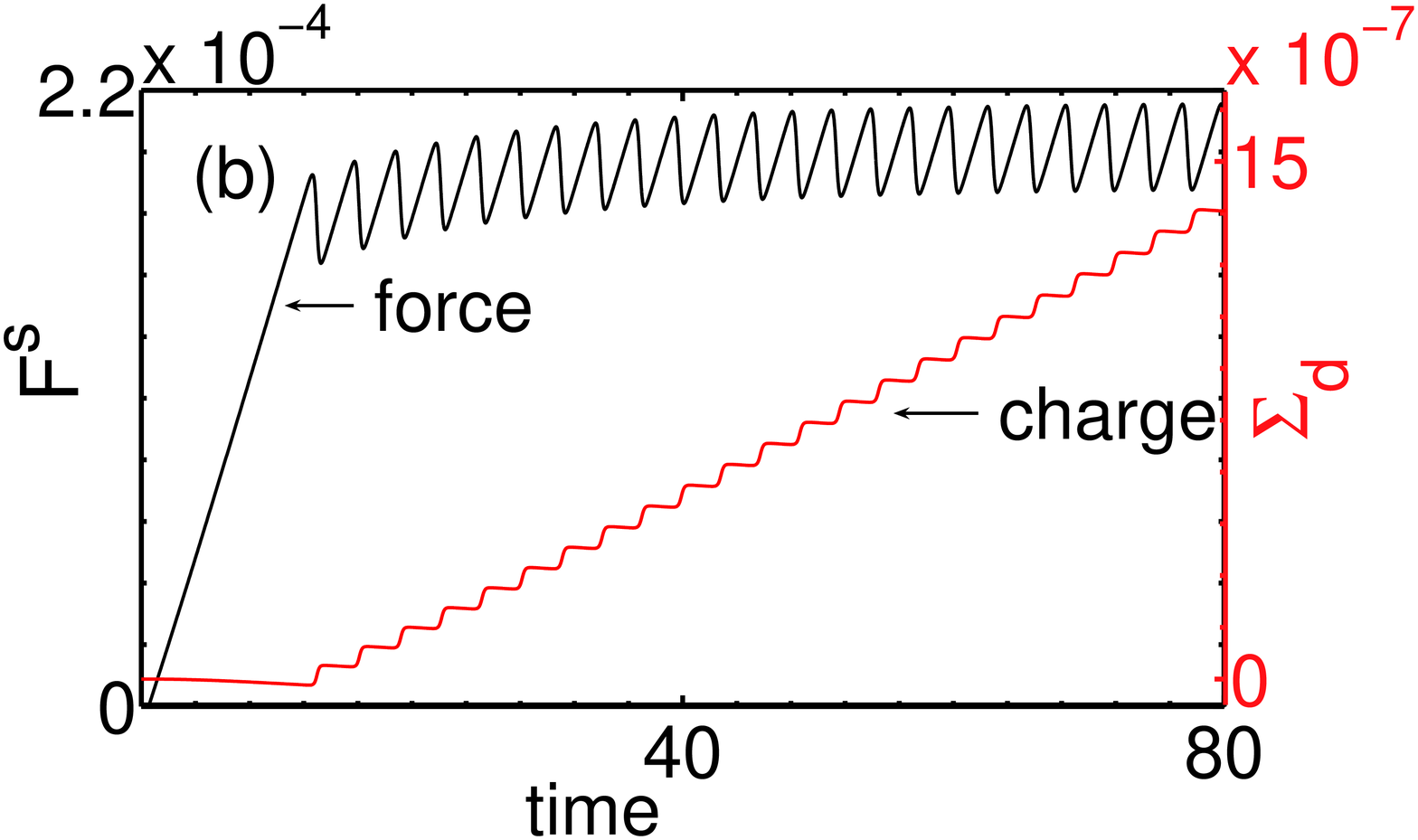}
}
\caption{(color online) (a) Plots of scaled force and cumulative charge transferred to the substrate for  $\bar{F}_n=3.5$. (b) Plots of scaled force and cumulative charge transferred to the substrate for  $\bar{F}_n=2.0$ and $\Sigma_m= 0.835$ ($\sigma_m = 5\times 10{-4} C/m^2$). 
}
\label{FPM1}
\end{figure}

In our model, as $A_n \propto F_n^{2/3}$, we expect the scale factor $\alpha$ to depend on the load. However, the  area changes by a factor 1.34  when $F_n$ is changed from $68$ to  $106 \,mN$ which translates to a small change (decrease) in $\alpha$ (or equivalently $\alpha_s$).  However, for higher load levels the value of the scale factor $\alpha^s$ does change significantly.  To see this, consider a  much higher load level, say  $\bar{F}_n = 3.5$.  The scale factor for this case is $\alpha_s$ is $2.34 \times 10^4$, which  falls well outside the mean value of $\alpha_s$ for the limited load range $\bar{F}_n=1.446-2.26$. A plot of the force-displacement curve is shown in Fig. \ref{FPM1}(a) for $\bar{F}_n=3.5$  along with the cumulative charge transferred to the substrate $\Sigma_d$.  Thus, {\it the lack of dependence of $\alpha$ on load} is clearly due to the limited range of loads studied in Ref. \cite{BP00}.  The plot also shows that  the model predicts {\it fewer} stick-events on an average with increase in $F_n$ for  the same scan length, a feature taken as a support for electronic origin of friction \cite{BP00}. The above  features emerge due to the separation of time scales of the stick (charging) phase and the slip (charge transfer) phase.

As stated earlier, at the charge density level of $10^8 {\rm charges}/mm^2$, the slip threshold is controlled by the friction threshold $A_{x,0}\tau_0 z^{3/2}$. However, when $\sigma$ is increased, the onset of the first force drop increases. A plot of the force-displacement curve is shown in Fig. \ref{FPM1}(a) for $\sigma = 5 \times 10^9 {\rm charges/mm^2}$ keeping $\bar{F}_n=2.0$. (The cumulative charge transferred to the substrate $\Sigma_d$ is also shown.) It is clear that the threshold for the first force drop as also the average asymptotic value of $F^s$ has increased from those for $\sigma = 10^8 {\rm charges/mm^2}$. (Compare with Fig. \ref{FPM} (a).)

\begin{figure}[!h]
\vbox{
\includegraphics[height=5.0cm,width=7.5cm]{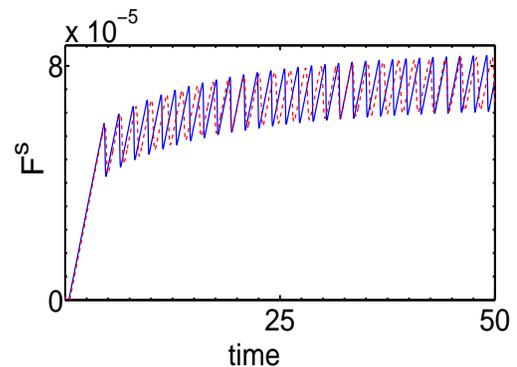}
}
\caption{(color online) (a) Plots of the scaled force -displacement curves for Model II (continuous blue) and Model I (dashed red).
}
\label{FPM3}
\end{figure}

As stated earlier, the results of the Model II are very similar to those of Model I. To see this, we first note that the equation for $X$ (Eq. \ref{X-eqn1}) for Model II differs from that of Model I [Eq. (\ref{X-eqn})] by a factor $Z^{3/2}$. However,  the near saturation value of $Z$ is almost entirely controlled  by Eq. (\ref{Z-eqn}). Thus,   the value of ${\eta}^{\prime}_{\|}$  where the instability occurs can be estimated by inserting the near saturation value of $Z$ in  $\bar{\eta}_{\|}={\eta}^{\prime}_{\|} Z^{3/2}$. This gives a value ${\eta}^{\prime}_{\|} \sim 4 \times 10^6$.  The rest of the parameters  are the same. A plot of  the force-displacement curve shown in Fig. \ref{FPM3} for $\bar{F}_n=2.0$. For the sake of comparison, we have also shown the force-displacement curve for Model I on the same plot as a dashed line. As can be seen, except for a phase factor, the two curves follow the same behavior. 

\begin{figure}[!h]
\vbox{
\includegraphics[height=5.5cm,width=7.5cm]{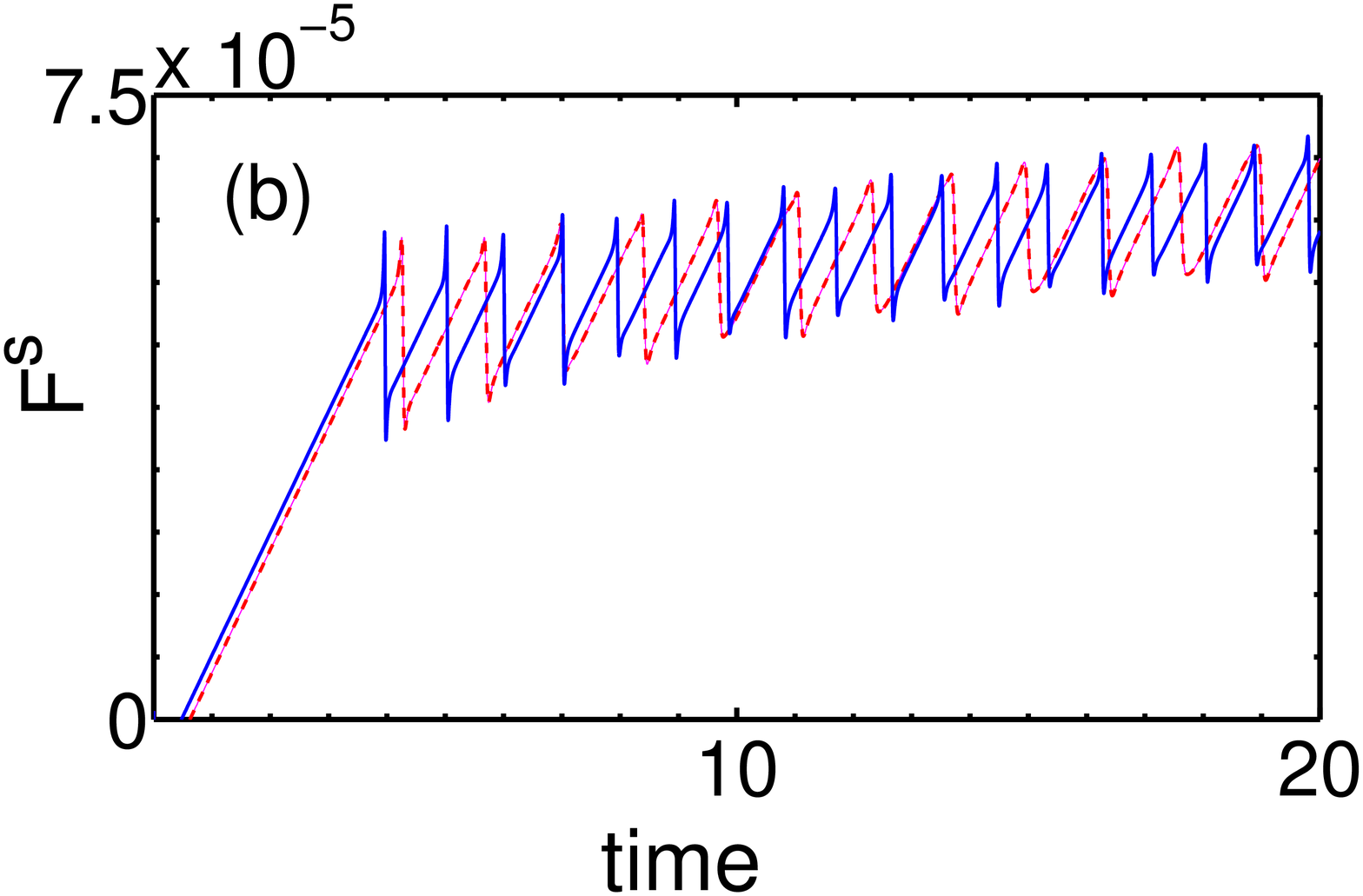}
}
\caption{(color online) (a) Plots of the scaled force -displacement curves for Model II (continuous blue) and Model I (dashed red) for $n=1.75$ and $q=2$.
}
\label{FPM4}
\end{figure}

The above calculations are for $n=q=1$. However, in general, they can be different from unity. Figure \ref{FPM4} shows the force-displacement curves for both Model I and II for $n=1.75$ and $q=2$. Again the difference between the two models is only in the phase. 

We now consider estimating the value of the scale factor $\alpha$. Since the force and charge transferred curves for a given scan length collapse onto a single curve {\it  only } in the asymptotic stick-slip  regime [Fig. \ref{FPM}(b)], we can use $\dot x \sim V_a$ valid on an average.  Using this in 
\begin{equation}
\int K_{\|} \dot x dt = \alpha \int \dot \sigma_d dt = \alpha \int (\dot x \sigma_t/D) dt
\end{equation} 
gives 
\begin{equation}
K_{\|} D \simeq \alpha \pi R z_{asy} \sigma_{asy},
\label{alpha}
\end{equation}
where $\sigma_{asy}$ is the mean value of $\sigma$ ($\sigma \rightarrow \sigma_m$) in the asymptotic stick-slip regime. Using $z_{asy}$ and $\sigma_m$  leads to $\alpha \sim 50$ $eV/{\buildrel _{\circ} \over {\mathrm{A}}}$ for $F_n =94$ $mN$. This value is high even after discounting the ideal nature of the model.

This prompted us to look for possible contributions that affect the scale factor $\alpha$. A cursory look at contact charge image in Fig. 2 of Ref. \cite{BP00} shows that the mean charging radius $R_c$ is surprisingly large, typically 10-20 times the contact radius $a$ with charge density $\sigma \sim 1.6- 2.4 \times 10^8 \,\, charges /mm^2$. While the authors make no comment on such a high value of charging radius, one possibility is that charge deposited spreads (possibly after the slip event) due to electro-static repulsion \cite{Lowell80,Lee94}. However, the model uses the natural choice that {\it charging area is equal to  the contact area} for that load. Thus, independent of the physical causes for such a large charging area, we should either use a larger contact area $\beta A_n$ with $\beta \sim (\frac{ R_c}{a})^2 \sim 100-400$ or  use higher charge density (by a factor $\beta$) keeping the contact area fixed. Using $\beta A_n$ in place of $A_n$, we get  $\alpha  \sim 0.5 - 0.125$ $eV/{\buildrel _{\circ} \over {\mathrm{A}}}$. This value is consistent with the reported value.
Finally, we note that the value of $\alpha$ is nearly the same for smooth sliding conditions   as  $\dot x = V_a$  is exact.

Another factor that affects contact charging is the surface roughness.  The influence of surface roughness on friction is well known in tribology.  Several models have been proposed to calculate the contribution arising from roughness \cite{GW66,Nayak71,Bush75}. The topic has been revisited recently with a view to accommodate the influence of a hierarchy of roughness scales \cite{Yang08}. However, for the load levels considered here, it is easy to show that this is likely to contribute about $10 \%$ \cite{Yang08}. Here, it may be relevant to point out that roughness in this case is likely to be dynamical due to the sliding over long intervals of time \cite{Berman98}. 

\section{Summary and Discussion}

In summary, we have developed a model for the stick-slip dynamics observed when the tip of the cantilever of a surface force apparatus is dragged on a substrate.  Equations of motion for the center of the contact area and penetration depth are set up by considering the tip as a single smooth asperity contacting with a smooth surface for which  contact mechanics is applicable. Further, the time dependent contributions arising from visco-elastic nature of the PMMA substrate and possible plastic deformation of the softer PMMA are included. The model exhibits stick-slip dynamics for a range of values of the parameters. To account for the contact charging at the area of contact, these equations are coupled to an equation of motion for the total contact charge developed on contact.    Charging occurs during the stick phase and transfer of charge to the substrate occurs during slip phase. 

The model captures several features of the experiment. First, {\it the correlation between the stick-slip events and charge deposited on the PMMA substrate is reproduced}. This is to be expected as the contribution from contact charging ( $\sim nN$) is orders of magnitude smaller than the static friction threshold ($\sim mN$) or even sliding frictional threshold. Second, for {\it higher load}, the model  {\it predicts  fewer events } for the same scan length. Third, lack of dependence of the scale factor $\alpha$ on normal load observed in experiments is traced to the {\it small range of values studied} in Ref. \cite{BP00}.  Indeed, Eq. (\ref{alpha}) shows that the scale factor $\alpha$ is inversely proportional to the area of contact for the load.  Fourth, in the model, contact charging is equal to the contact area for the load.  If this relation is used,  the magnitude of the scale factor turns out to be nearly 50 times the experimental value.  However, as discussed, the experimental charging  area  is  $\beta \sim 100 - 400$ times the theoretically calculated contact area for that load (see Fig. 2 of Ref. \cite{BP00}). Using this value in place of the contact area in Eq. (\ref{alpha}) gives $\alpha \sim 0.5 -0.125 $ $eV/{\buildrel _{\circ} \over {\mathrm{A}}}$.  This value is close to the measured value.  Since the lack of dependence of $\alpha$ on load together with the value of $\alpha$ has been interpreted to mean that friction is controlled by contact charging, it is necessary to check the lack of dependence of $\alpha$ over larger range of normal loads before the suggestion can be taken seriously. Further, the value of $\alpha$ should be nearly same for smooth sliding conditions as well.  All these features follow naturally due to the separation of time scales of stick and slip events. Stick-slip events are not affected by {\it contact charging} for the charge levels in Ref. \cite{BP00}.   Instead plastic deformation of the interface material is the cause of slip.   At a mathematical level the instability is due to a competition between the visco-elastic and plastic deformation time scales with the applied time scale. As most experimental features are captured, the model provides an alternate explanation for the results.  Lastly, the fact that sliding friction is lower than static friction comes out naturally from the equations of motion.

Finally, the model should be applicable to other stick-slip phase observed in frictional sliding experiments where plastic deformation is significant, including stick-slip observed during scratching of polymer sheets \cite{Li96,VH00,HV00}.\\

\centerline{ {\bf ACKNOWLEDGMENTS}}
G. A would like to acknowledge financial support from BRNS grant 2007/36/62-BRNS and Raja Ramanna Fellowship scheme.


\begin{thebibliography}{99}
\bibitem{Note}  A better way of dealing with plastic deformation in the $x$ (or in the $z$) direction would be to write an equation of plastic strain rate and couple it to the equation of motion for $x$ (or for $z$). This is not attempted here since the equations of motion are already complicated.
\end{thebibliography}
\end{document}